# Scanning Acoustic Microscopy for Quantifying Bubble Evolution in Alkaline Water Electrolyzers

Zehua Dou [1, 7, *], Hannes Rox [2], Zyzi Ramos [3], Robert Baumann [4], Rachappa Ravishankar [8], Peter Czurratis [3], Xuegeng Yang [2], Andrés Fabian Lasagni [4, 5], Kerstin Eckert [2, 6, 7], Juergen Czarske [1] and David Weik [1]

[1] Faculty of Electrical and Computer Engineering, Laboratory of Measurement and Sensor System Technique, Technische Universität Dresden, Helmholtzstraße 18, 01069 Dresden, Germany

[2] Institute of Fluid Dynamics, Helmholtz-Zentrum Dresden-Rossendorf, Bautzner Landstraße 400, 01328 Dresden, Germany

[3] PVA TePla Analytical Systems GmbH, Deutschordenstraße 38, 73463 Westhausen, Germany

[4] Institute of Manufacturing Science and Engineering, Technische Universität Dresden, George-Baehr-Straße 3c, 01069 Dresden, Germany

[5] Fraunhofer Institute for Material and Beam Technology IWS, Winterbergstraße 28, 01277 Dresden, Germany

[6] Institute of Process Engineering and Environmental Technology, Technische Universität Dresden, Helmholtzstraße 14, 01069 Dresden, Germany

[7] Hydrogen Lab, School of Engineering, Technische Universität Dresden, Helmholtzstraße 14, 01062 Dresden, Germany

[8] Institute for Emerging Electronic Technologies, Leibniz Institute for Solid State and Materials Research Dresden, Helmholtzstraße 20, 01069 Dresden, Germany

[*] zehua.dou@tu-dresden.de

## Abstract

Improved understanding of gas/liquid transport in electrochemical gas-evolving systems is increasingly demanded for optimizing device performance. However, high-resolution measurement techniques for *in-situ* imaging remain limited. This work demonstrates the use of volumetric scanning acoustic microscopy (SAM) for quantifying hydrogen bubble evolution in porous nickel electrodes in a customized alkaline water electrolysis cell. By using high-frequency focused ultrasound, SAM enables volumetric imaging with high spatial resolution in the range of tens of micrometers. This allows the distribution of gas bubbles within the complex 3D architecture of porous electrodes to be resolved. Digital image processing methods are used to segment and quantify the gas content in the electrode. Thus, non-destructive SAM imaging is demonstrated to be an accessible and scalable analytical tool for the quantitative investigation of bubble evolution in operando electrochemical environments. Here, a solid foundation is established for future studies aimed at optimizing bubble dynamics and cell design under practically relevant operating conditions, ultimately contributing to higher electrolysis efficiencies.

## I. Introduction

Low temperature water electrolysis in alkaline water electrolyzers (AWEs) and proton exchange membrane water electrolyzers (PEMWEs) are key technologies for producing green hydrogen at large scale [1], [2]. However, besides substantial investment costs [3], both technologies still display an insufficient energy efficiency. In PEMWEs electrolysers, the efficiency at the cell level is 60–80% [4], whereas the efficiency of AWE is around 70–80% [5], based on the higher heating value. Efficiency at the cell level is influenced by several factors, such as kinetic inhibition at the electrodes, and transport resistance of the reactants and products, giving rise to specific overpotentials. It is meanwhile accepted that beside the activation overpotentials, the bubble-induced overvoltages are a main issue [4], [6]. Addressing this challenge requires advanced imaging techniques that are capable of quantifying bubble dynamics in realistic practical operating environment and conditions, thereby enriching insights into developing AWEs with optimal performance [7], [8].

Up to date, optical imaging, synchrotron X-ray and neutron radiography and computed tomography (CT) systems have been extensively applied to investigate fluid transport in electrolyzers [9]. Direct visualizations with high speed cameras, on one hand, present a straightforward approach for real-time quantifications of e.g., electrode coverage due to bubble evolution and bubble size distribution, in opened electrolysis systems and transparently designed test AWEs [10]–[12]. However, optical imaging systems are incapable of investigating bubble dynamics inside optically opaque porous electrodes. Synchrotron X-rays and neutrons, on the other hand, are able to deeply penetrate electrolyzer components. Advanced CT instrumentations based on modern high flux beamlines show great potential to achieve 3D imaging in real-time with micrometer level spatial resolutions [13], [14]. Taking these advantages, synchrotron X-ray and neutron imaging have been extensively used in PEMWE related studies [15]–[19], and a few investigations of AWEs were reported as well [20]–[22]. However, these imaging systems generally exhibit insufficient contrast between gas, liquid, and metal, which limits complete visualization of liquid transport in technically relevant electrodes. In particular, X-rays are highly attenuated in metallic materials, e.g., Titanium (Ti) based porous transport layers (PTL) of PEMWEs, resulting in poor gas/liquid contrast [17], [18]. On the contrary, neutrons are strongly attenuated by water, which means that the metallic electrodes cannot be distinguished from the gas bubbles generated [23], [24]. Such limitations are likely to persist in AWEs related studies, as the attenuation coefficients of X-rays and neutrons in AWE electrode materials e.g., nickel (Ni) are similar to the aforementioned values of the Ti-PTL in PEMWEs [25], [26]. Although the contrast issue can be solved by introducing contrast agents or combined X-ray and neutron imaging schemes [23], [27], the complicated and costly instrumentations, as well as the limited access to beamlines hinder the laboratory-based research work flows. Nonetheless, the CT systems face inherent trade-off among the imaging volume, temporal resolution and imaging quality, hindering their transfer to large systems.

Recently, ultrasonic imaging techniques have opened new horizons for studying fluid transport in electrolyzers using cost-effective instrumentation. Moreover, the contrast in ultrasonic imaging is determined by the mismatch of the acoustic impedances of different materials involved. This enables clear differentiations between metallic electrodes, liquid electrolyte, and gas bubbles based on their drastically different acoustic impedances. Pulse-echo ultrasonic imaging of *operando* PEMWEs with an active area of 9 $cm^2$ using a 5 MHz linear ultrasonic transducer array was presented in [28], whereby the flow field, the porous electrode, and the gas content in the flow channels could be clearly distinguished. However, the spatial resolution achieved was only 1 mm, which prevents the detailed visualizations of bubble dynamics within the porous electrode. The limited spatial resolution is due to the fact that the low-frequency linear transducer array is unable to focus the ultrasound correctly.

Aiming for resolving bubble dynamics at pore scale of $O(100\ \mu m)$, high frequency focused ultrasound beams with a diameter of $O(10\ \mu m)$ must be employed. One implementation of such an instrumentation is known as scanning acoustic microscopy (SAM) that is able to provide a spatial resolution up to sub-micrometer in opaque materials. Taking these advantages, SAM has found extensive applications in non-destructive evaluations for material researches and the semiconductor industry [29], [30], characterizations of biological tissues [31], and it has nowadays emerged as an efficient approach to study the degradations of Lithium-ion batteries [32]. Such a powerful tool thus shows tremendous potential for applications in electrolyzers, and is a valuable complement to the costly and complicated X-ray and neutron-based imaging methods.

In this work, we experimentally validate the feasibility of quantifying the bubble evolution in AWEs using volumetric SAM imaging. To showcase the general applicability of this imaging technique, we investigate Ni woven mesh and open cell Ni foam electrodes that are representative for a wide range of technically relevant AWE electrodes. Volumetric images of the electrodes during the hydrogen evolution reaction (HER) in a customized AWE are acquired with a leading-edge SAM system, providing a lateral resolution of approx. 70 µm and an axial resolution of approx. 10 µm. Two tailored image segmentation approaches are employed to quantify gas volumes in each electrode type, providing direct measurements of bubble accumulation and its spatial evolution under electrochemical operation.

## 2. Materials and methods

### 2.1 The assembly and operation of the test cell

#### 2.1.1 The investigated porous electrodes

In this work, two different types of electrodes were investigated, namely Ni woven mesh and open cell Ni foam. The investigated electrodes possessed a dimension of 300 mm$^2$ (30 × 10 mm$^2$). The thicknesses of the investigated woven mesh and open cell foam electrodes were 140 µm and 1.5 mm, respectively. The woven mesh, **Fig. 1 (a),** featured a wire diameter of 140 µm and a pore size of 500 µm while the open cell foam, **Fig. 1 (b)**, displayed a porosity of approx. 94% and a pore diameter of approx. 200 µm.

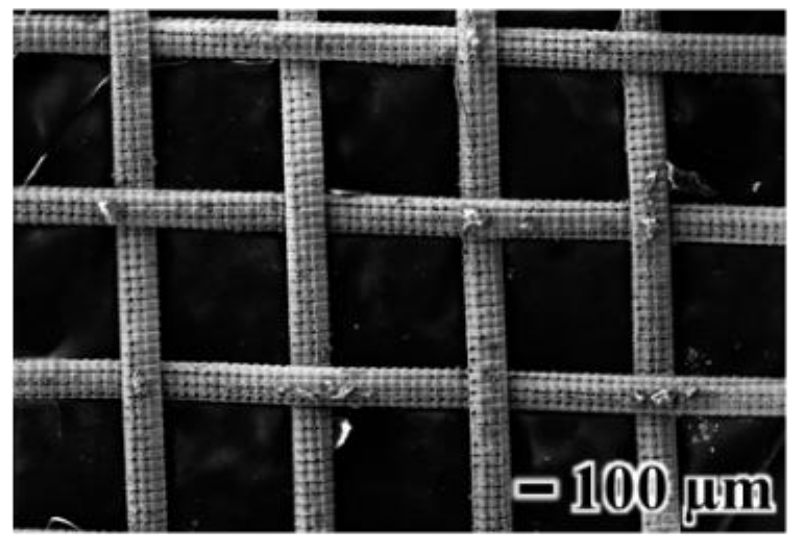

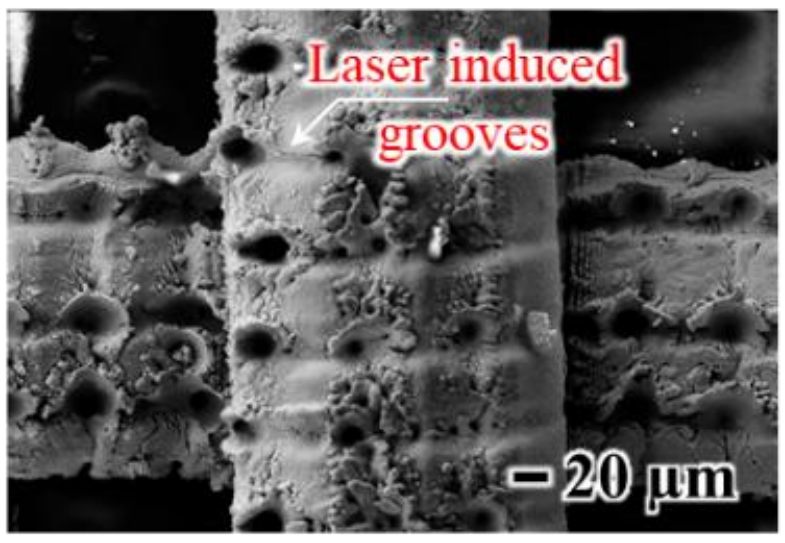

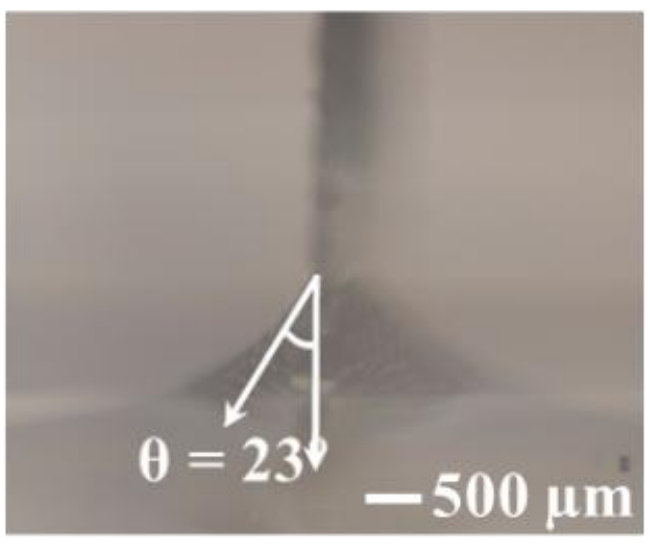

(a) The laser textured Ni woven mesh (LNM)

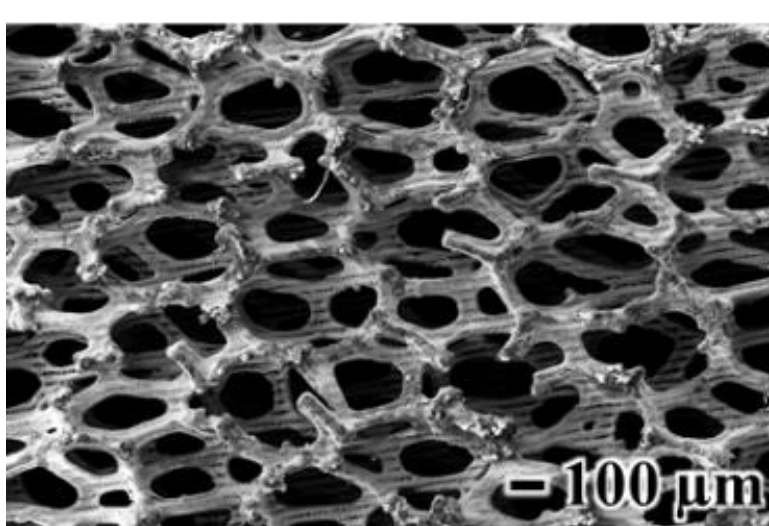

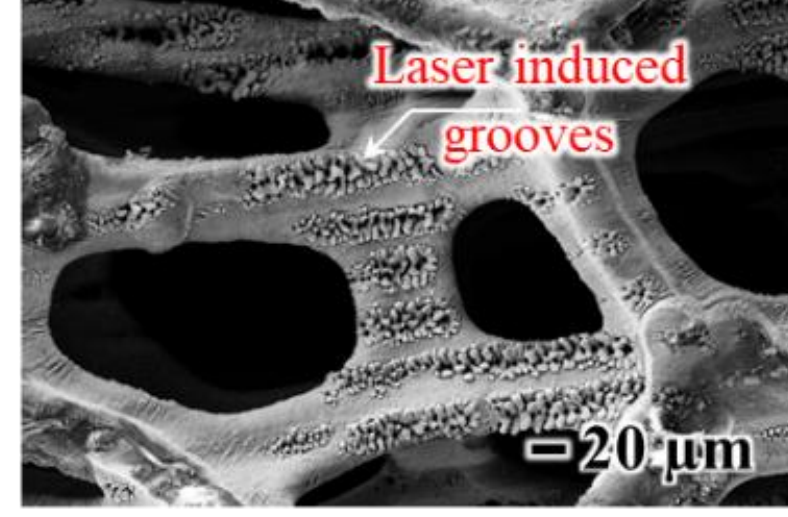

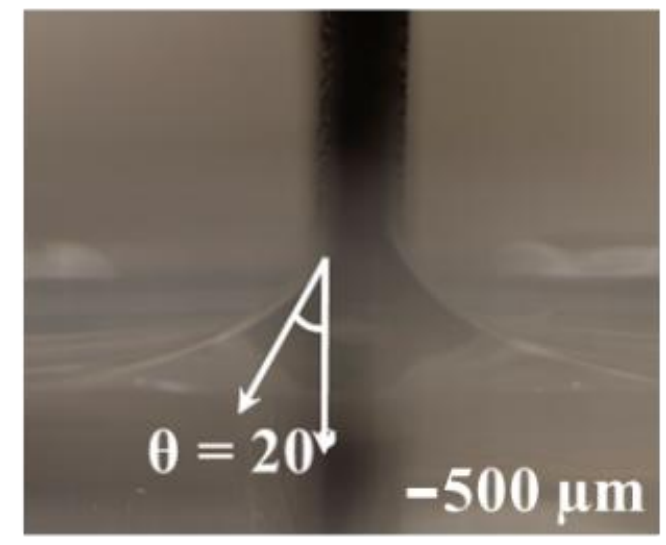

(b) The laser textured Ni open cell foam (LNF)

**Fig. 1. SEM images and apparent contact angles of (a) the laser textured Ni woven mesh and (b) the laser textured open cell Ni foam (LNF).**

In addition, the surfaces of the two electrodes were textured with direct laser writing (DLW) [33]. The laser-induced grooves are clearly visible in the high-resolution SEM images, as shown in the middle column of **Fig. 1 (a) & (b)**. For the sake of brevity, the laser-textured Ni mesh and the open-cell Ni foam electrodes will be referred to as LNM and LNF in the rest of this article. The detailed process and parameters of DLW are given in **supplementary information (SI)**.

A significant advantage of DLW is its ability to achieve wetting without relying on specific electrocatalytic nanostructure [34]. This is confirmed by contact angle measurements using the Wilhelmy plate method, which yielded apparent contact angles of 23 ° for the LNM and 20 ° for the LNF, as shown in the right-hand column of **Fig. 1 (a) & (b)**. In other words, both electrodes exhibited a wicking state in the electrolyte, which is also reported in [35]. The detailed characterization procedures of the surface morphology and contact angles of the electrodes are described in **SI**.

2.1.2 Design and operations of the test AWE

To investigate bubble evolution in the two above described electrodes in an environment closer to practical application, a miniaturised zero-gap AWE was developed in-house using 3D printing (Objet30 Prime V5, Stratasys, USA) as shown in **Fig. 2 (a) & (b)**. In this design, the single straight flow channel allowed the electrolyte to flow parallel to the electrode surface [as marked by the blue shade in **Fig. 2 (a)**]. The LNM and LNF were used as the working electrodes (WE), and thus, cathode, for the hydrogen evolution reaction (HER). A ring-shaped Ni woven mesh served as the counter electrode at the anode side [see **Fig.2 (c)**]. To measure the WE's potential during the galvanostatic measurements, a Platinum (Pt) wire was used as a pseudo-reference electrode (RE) at the cathode side due to space constraints, as suggested in [36]. A separator was placed between the anode and cathode to avoid short circuit and gas crossover. In the cathodic flow field, the channel width of 11.5 mm partially overlaid with the investigated electrodes in the cathode, and the rest section of the electrodes out of the flow channel was thus analogue to the case of under a rib [18], [27]. To ensure proper assembly of the electrodes, the cathodic and anodic flow fields of the test AWE were sealed by PMMA plates with a diameter of 30 mm and a thickness of 4 mm.

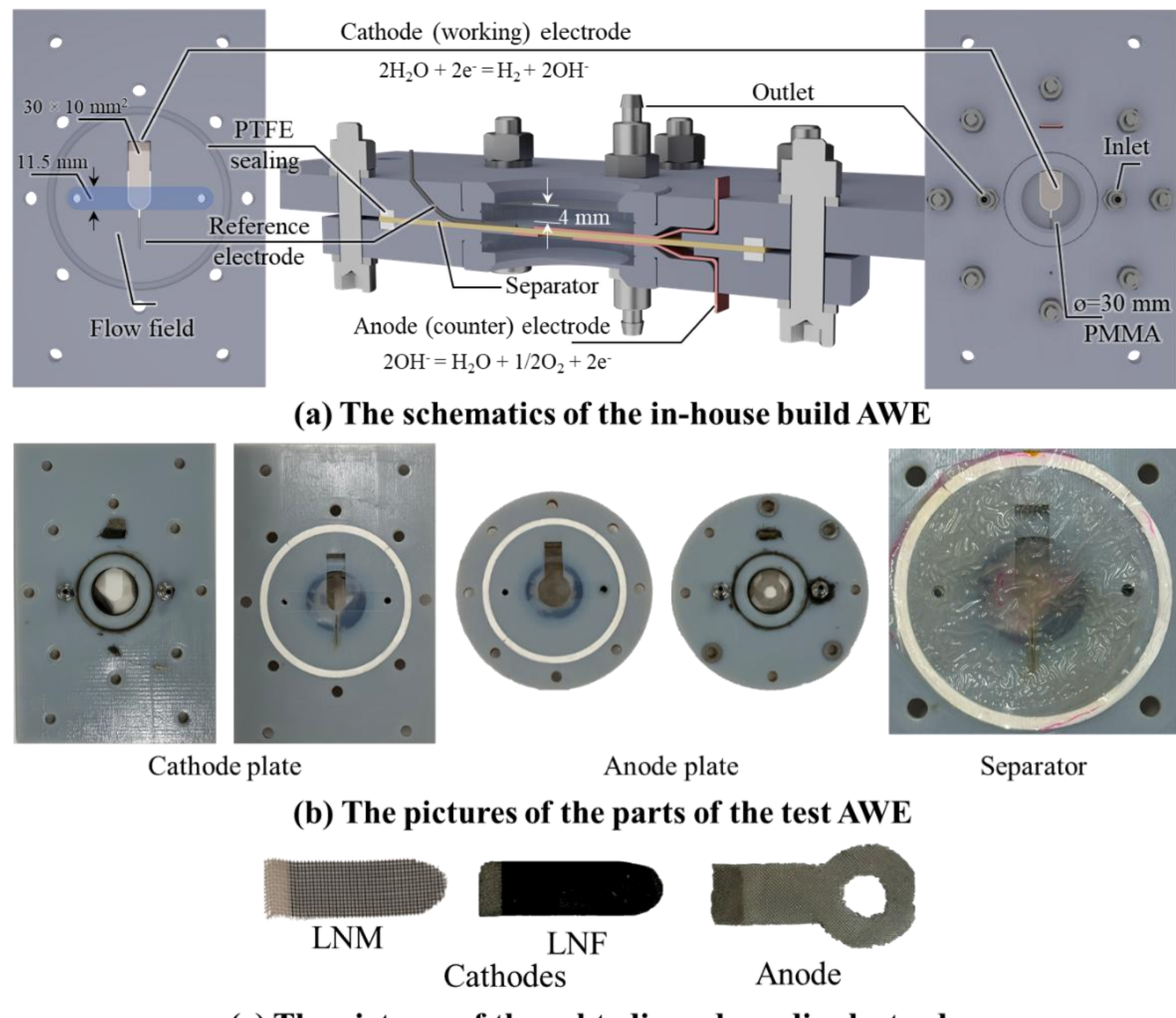

**Fig. 2. Design of the in-house developed test alkaline water electrolyzer (AWE):** (a) The schematics of the test AWE, in which the key dimensions are given. (b) The images of the parts of the test AWE. (c) The images of the investigated LNM and LNF and the counter electrode.

The AWE was operated at room temperature, under atmospheric pressure, and supplied with 8 wt% KOH solution as electrolyte to ensure the chemical integrity of the 3D printed parts during experiments. An electrochemical workstation (CHI660I, CH Instruments, USA) was used for galvanostatic electrolysis.

## 2.2. Volumetric SAM imaging for AWE

### 2.2.1 Working principle of SAM imaging

SAM imaging in pulse-echo mode is schematically illustrated in **Fig. 3 (a)**. In this technique, a high-frequency ultrasonic transducer emits short acoustic pulses into the sample. When the ultrasonic pulse propagates to an interface of two materials with different acoustic impedances e.g., bubble and electrolyte, it is partially reflected or scattered, which is then acquired by the transducer.

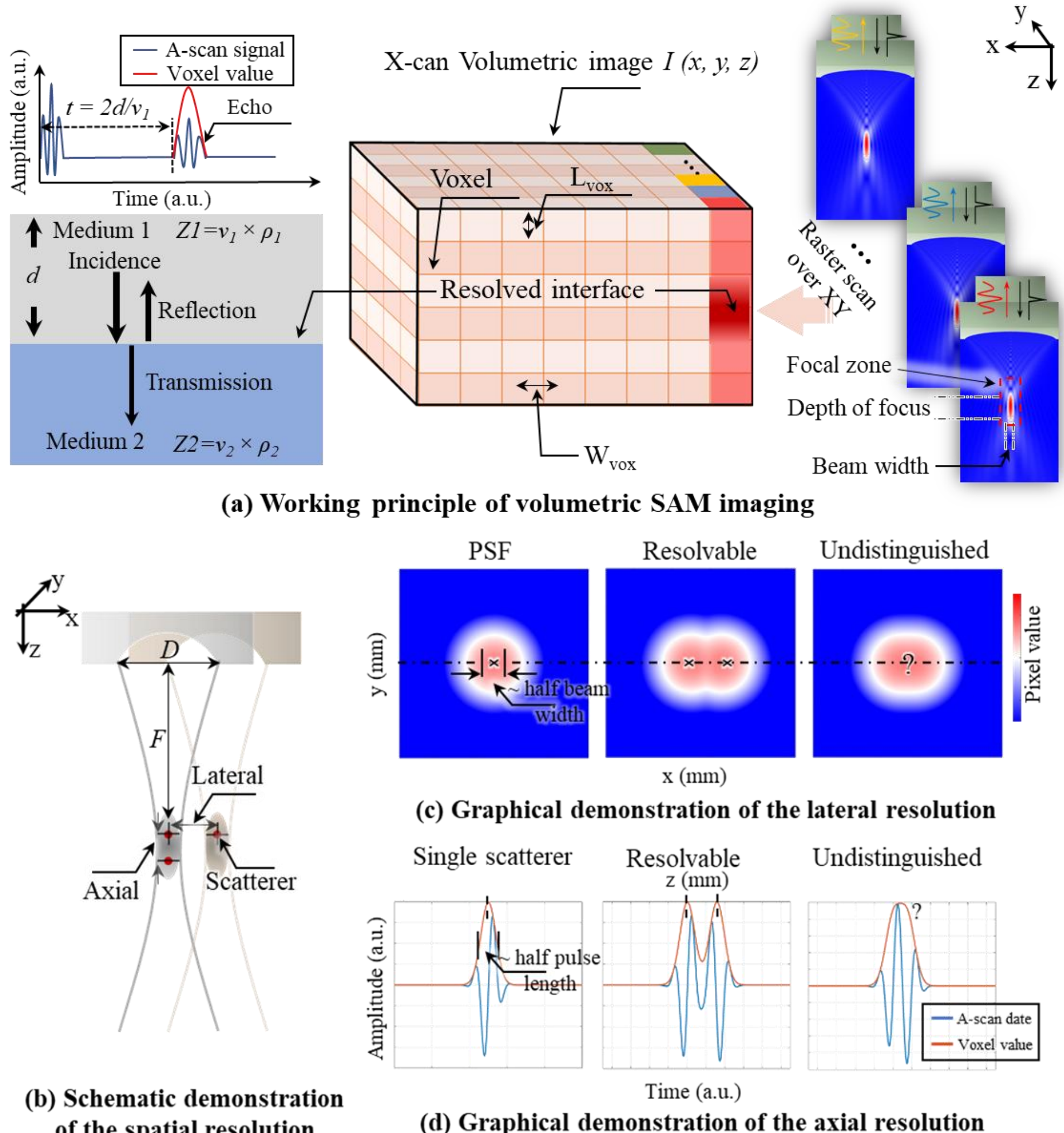

**Fig. 3. Fundamentals of volumetric SAM imaging.** (a) The working principle of SAM imaging. (b) The definition of spatial resolution in SAM imaging. (c) Graphical demonstration of the axial resolution. (d) Graphical demonstration of the lateral resolution.

The strength of each reflection depends on the acoustic impedance mismatch at the interface. Specifically, the acoustic impedance, $Z = \rho c$, of a material is defined as the product of its density, $\rho$, and the speed of sound, $c$. The reflection coefficient of acoustic energy, R, between two media with impedances $Z_1$ and $Z_2$ is given in **(1)**.

$$R = \left( {Z_1 - Z_2}/{Z_1 + Z_2} \right)^2 \tag{1}$$

In **(1)**, $Z_1$ and $Z_2$ represents the acoustic impedance values of the two mediums forming an interface. **Eq. (1)** also indicates that a greater mismatch in acoustic impedance results in stronger acoustic reflection.

For a clear perception of the SAM imaging in the experiments, the acoustic properties of the materials in the test cell are summarized in **Table I.** It is found that the gas bubbles with very low acoustic impedances embedded in water or nickel electrodes, possessing a much higher acoustic impedance, produce strong echoes, making ultrasound imaging particularly suitable for investigating bubble evolution.

**TABLE I. The Acoustic Properties of the Different Components of the Test AWE**

| Materials | In AWE | Density ($kg/m^3$) | Speed of sound (m/s) | Acoustic impedance ($10^6$ $rayl/m^2$) [b] |
|---|---|---|---|---|
| DI water [37] | Coupling agent | 1000 | 1450 | 1.45 |
| PMMA [38] | Flow field plate (electrode region) | 1180 | 2730 | 3.19 |
| SLA resin [a] | Flow field plate | 1100 - 1300 | 2350 | 2.59 – 3.06 |
| Epoxy [39] | Jointing PMMA and SLA resin | 1070 | ~ 2000 | 2.14 |
| 8 wt% KOH solution [a] | Electrolyte | 1069 | 1580 | 1.69 |
| Ni [c] [40] | Electrode | 8900 | 6040 | 53.76 |
| Air [37] | Bubbles [d] | 1.293 | 340 | $4.4 \times 10^{-4}$ |
| $H_2$ [41] | Bubbles | 0.083 | 1310 | $1.1 \times 10^{-4}$ |

[a] Experimentally determined in this work.

[b] Reflection coefficients at different phase interfaces can be estimated from the acoustic impedance mismatch. The unit of acoustic impedance in the MKS (meter, kilogram, second) system is rayl per square meter ($Rayl/m^2$) that is pascal-second per cubic meter in SI unit system.

[c] Note that these listed values are for pure Ni but not the porous Ni electrodes.

[d] Bubbles in the AWE can either be air bubbles that are initially trapped into the electrodes, or $H_2$ bubbles produced by electrolysis.

Aiming for high resolution imaging down to the pore scale, SAM employs a spherically focused transducer i.e., a transducer with a diameter D and a certain radius of curvature that concentrates the ultrasound beam into a narrow focal region within the sample, as shown in the right diagram of **Fig. 3 (a)**. This focused beam increases the local acoustic energy and limits the width of the interaction zone, which contributes significantly to improving lateral resolution.

The echo received by the transducer is captured as a 1D time-domain signal, known as an A-scan, in which the wave packet from each scatterer/reflector within the focal zone appears at the transit time corresponding to the round-trip wave propagation. Based on the known speed of sound in the medium, the A-scan provides depth-resolved information along the axial ($z$) direction by assigning the intensity of each sampled data point in the A-scan signal to the voxel at the corresponding depth, as shown in **Fig. 3 (a)**. Repeating this A-scan process across the horizontal plane ($x$-$y$) via raster scanning allows for the construction of a full 3D volumetric image, which is termed as the X-scan mode. The result is a stack of depth-resolved signals that form a volumetric dataset, illustrated schematically in the centre of **Fig. 3 (a)**.

The voxel size in the axial direction ($L_{vox}$) is defined in Eq. (2), in which fs is the sampling rate of the analog to digital converter (ADC), and c is the speed of sound in the medium.

$$L_{vox} = \frac{c}{2 \cdot fs} \tag{2}$$

While, the effective imaging depth of the X-scan mode is determined by axial range of the focal zone provided by the transducer, which is referred to the depth of focus (DoF). Within the DoF, the focused beam remains sufficiently narrow, maintaining high resolution. Outside this focal zone, the beam diverges and the spatial resolution significantly degrades. Therefore, for volumetric imaging of thicker samples, multiple X-scans at incremented axial positions may be required to fully capture structures beyond the DoF of a single scan.

Furthermore, the voxel size in the lateral direction ($W_{vox}$) is defined by the scanning step size. It is however worth to emphasise that the voxel size does not directly correspond to the actual spatial resolution. The former represents the density of spatial sampling, while the latter refers to the minimum distance where two distinct objects can be resolved, as depicted in **Fig. 3 (b)**. More precisely, the lateral resolution ($r_l$), on the horizontal plane, is approximately the half width of the ultrasound beam at the focal zone, as shown in **Fig. 3 (c)**, which is known as the diffraction limit. The formula is given in **(3)** where λ is the wavelength, F is the focal length, and D is the aperture diameter of the transducer [42].

$$r_l \approx \frac{\lambda \cdot F}{D} \tag{3}$$

Similarly, the axial resolution ($r_a$), along z direction, is approximately half of the ultrasonic pulse spatial length in the medium, as illustrated in **Fig. 3 (d).** It is given in **(4)** where $\tau$ and $v$ are the pulse duration and speed of sound in the medium, respectively [43].

$$r_a \approx \frac{v \cdot \tau}{2} \tag{4}$$

2.2.2 Volumetric SAM image acquisition scheme for the test AWE

For investigating the bubble dynamics associated with the HER, the entire cathode flow field of the test AWE covering an area of 30 mm × 30 mm was designated as the region of interest (RoI) for SAM imaging, as illustrated in **Fig. 4 (a)**.

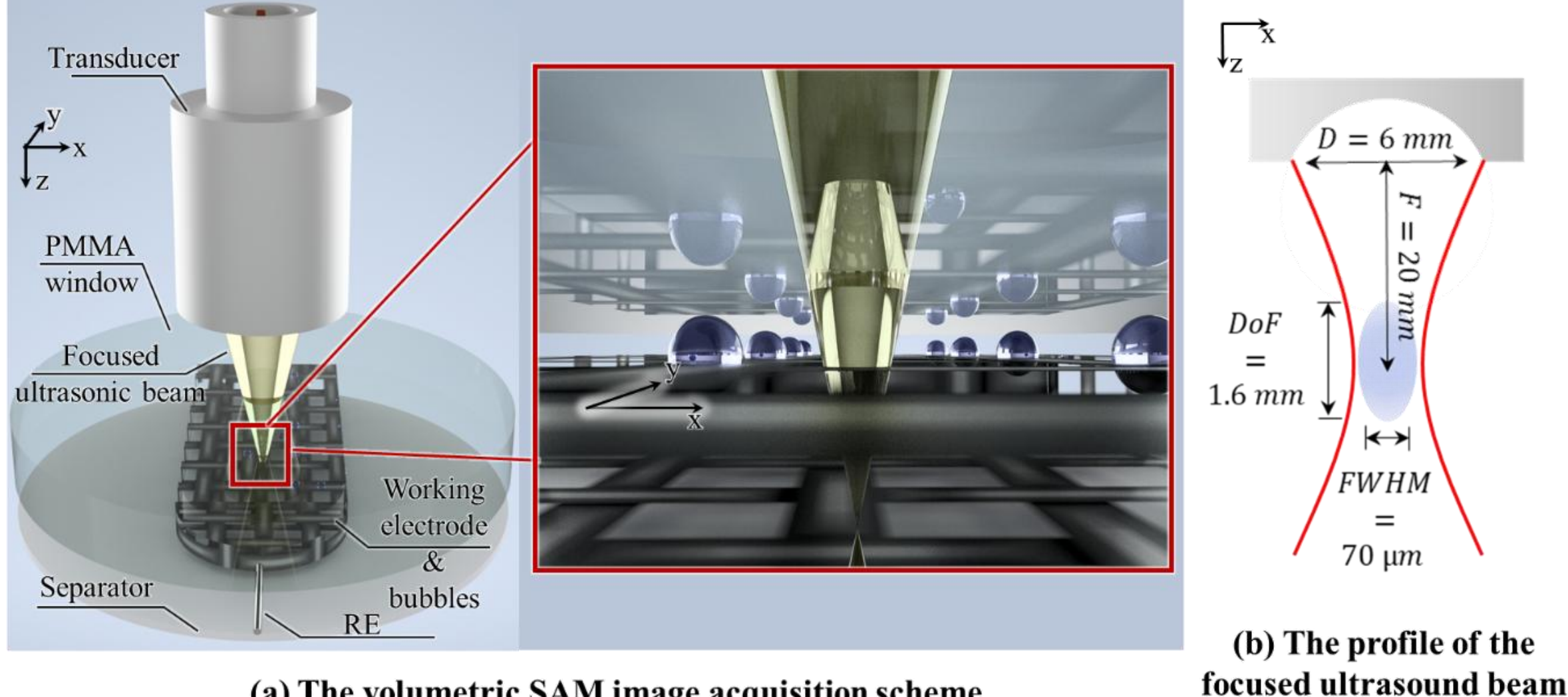

**Fig. 4. The *Operando* volumetric SAM image acquisition scheme.** (a) The schematic diagram showing SAM imaging of the cathode flow field of the test cell where the focal zone of the transducer is well positioned inside the electrode. (b) Profile of the focused ultrasound beam in the AWE environment.

In the experiments, volumetric images of the test AWE during electrolysis were acquired by a commercial SAM system (SAM 501 HD², PVA TePla AG, Westhausen, Germany). This system is capable of providing a lateral scanning precision of 100 nm. A more detailed technical specification of the system can be found elsewhere [44]. To ensure effective acoustic coupling during SAM imaging, the AWE was fully immersed in an aquarium filled with Milli-Q deionized (DI) water. As the flow field of the AWE cell was completely sealed, the electrolysis remained unaffected by the water immersion. An ultrasonic transducer with a mid-frequency of 75 MHz, which is capable of generating 13 ns pulses, was used for the experiments. According to the manufacturer's specifications, this transducer is able to detect rigid objects as small as 6.7 µm. The transducer is also expected to detect even smaller gas bubbles, due to the strong acoustic impedance mismatch between gas bubble and liquid electrolyte. The selected transducer was equipped with a spherical focal lens with a diameter of 6 mm and a focal length of 20 mm. Based on the values given in **Table I**, the profile of the focused beam in the AWE environment can be calculated, as shown in **Fig. 4 (b)** where the depth of focus (DoF) was 1.6 mm, and the half beam width at focus was 70 µm.

Since the DoF of 1.6 mm fully covered the electrode thickness of 1.5 mm, volumetric images of fluid transport within the investigated electrodes were able to be acquired with single X-scans at a constant height. For this, the axial position (z-axis) of the transducer was carefully adjusted, so that the focal zone of the ultrasound beam was centralized in the electrode, as illustrated in the magnified view in **Fig. 4 (a)**. The transducer was then scanned pointwise across the horizontal (*x*-*y*) plane over the RoI.

As raster scanning is the most time-consuming part of SAM image acquisition, a trade-off between imaging quality and acquisition speed must be carefully considered. Thus, in the experiments, a lateral scanning step size of 20 µm, finer than the lateral resolution limit of 70 µm, was selected to ensure spatial oversampling and enable robust image reconstruction. Using this configuration, each X-scan took approx. 180 seconds to acquire all the A-scan signals required to construct a single volumetric frame over the RoI area of 900 mm² (30 mm × 30 mm). The image reconstruction was directly accomplished by the build-in software of the commercial SAM system. Each volumetric frame generated heavy data exceeding 1 GB, reflecting the high spatial resolution and data density of the imaging technique.

**2.3 Characterizations of volumetric SAM Imaging performance**

With the working principle and experimental procedure of volumetric SAM imaging established, we next evaluate its imaging performance in the test AWE environment. In **Fig. 5 (a)**, an exemplary A-scan recorded from the laser textured nickel foam (LNF, Section 2.1.1) is shown, representing the signal quality for volumetric image construction. Specifically, the segment of the waveform between 18.20 µs and 21.16 µs corresponds to the round-trip ultrasound propagation time in the 4 mm thick PMMA plate. This can be verified using the known speed of sound in PMMA listed in **Table I** to calculate the thickness of the PMMA plate, $d_{pmma}$:

$$d_{PMMA} = {c_{PMMA} \cdot \Delta t}/{2} = 2730\ {m}/{s} \times {(21.16 - 18.2)\ \mu s}/{2} = 4\ mm \qquad (5).$$

The waveform after 21.16 µs thus corresponds to the depth range occupied by the electrode that was positioned right behind the rear surface of the PMMA plate. Within this region, the blue-shaded window, lasting 1500 ns and exhibiting a high signal-to-noise ratio (SNR), was selected for high-quality image reconstruction. This time window translates into an equivalent penetration depth of

$$d = {v_{KOH} \cdot \Delta t}/{2} = 1580\ {m}/{s} \times {1500\ ns}/{2} = 1.185\ mm \qquad (6),$$

in the KOH electrolyte saturated electrode material. This penetration depth is sufficient for investigating fluid transport in the woven mesh type electrodes with typical thicknesses of several hundred micrometres, as well as in the open cell foam electrodes that are thinner than this value. However, it was not adequate for the particular LNF addressed in this study, which possessed a thickness of 1.5 mm. The reduced penetration in the LNF was mainly due to strong acoustic scattering arising from its stochastic porous structure, where incident ultrasonic waves are partially diffused, resulting in a significant loss of coherent signal amplitude and consequently lower SNR, particularly at greater depths. It is important to emphasize that this limited penetration was not a fundamental limitation of the SAM technique itself, but rather a result of the technical implementation in this proof-of-concept study. Enhanced SNRs and hence greater penetration depths can be achieved by e.g., transducers with larger apertures and higher transduction efficiencies.

For image reconstruction, all the sampling points within the selected time window were converted into grey scale values, with their intensities referenced to that of the echo signal from the front surface of the PMMA plate, as marked by the green line in **Fig. 5 (a)**. This referencing served as an adaptive calibration to ensure a consistent grey scale across all the acquired frames, compensating for any potential variations in transducer performance or experimental conditions. The reconstructed volumetric images of the Nickel Mesh (LNM) and Nickel Foam (LNF) electrodes are shown in **Fig. 5 (b) & (c)**, respectively, in which the electrodes and structures of the test cell are clearly resolved.

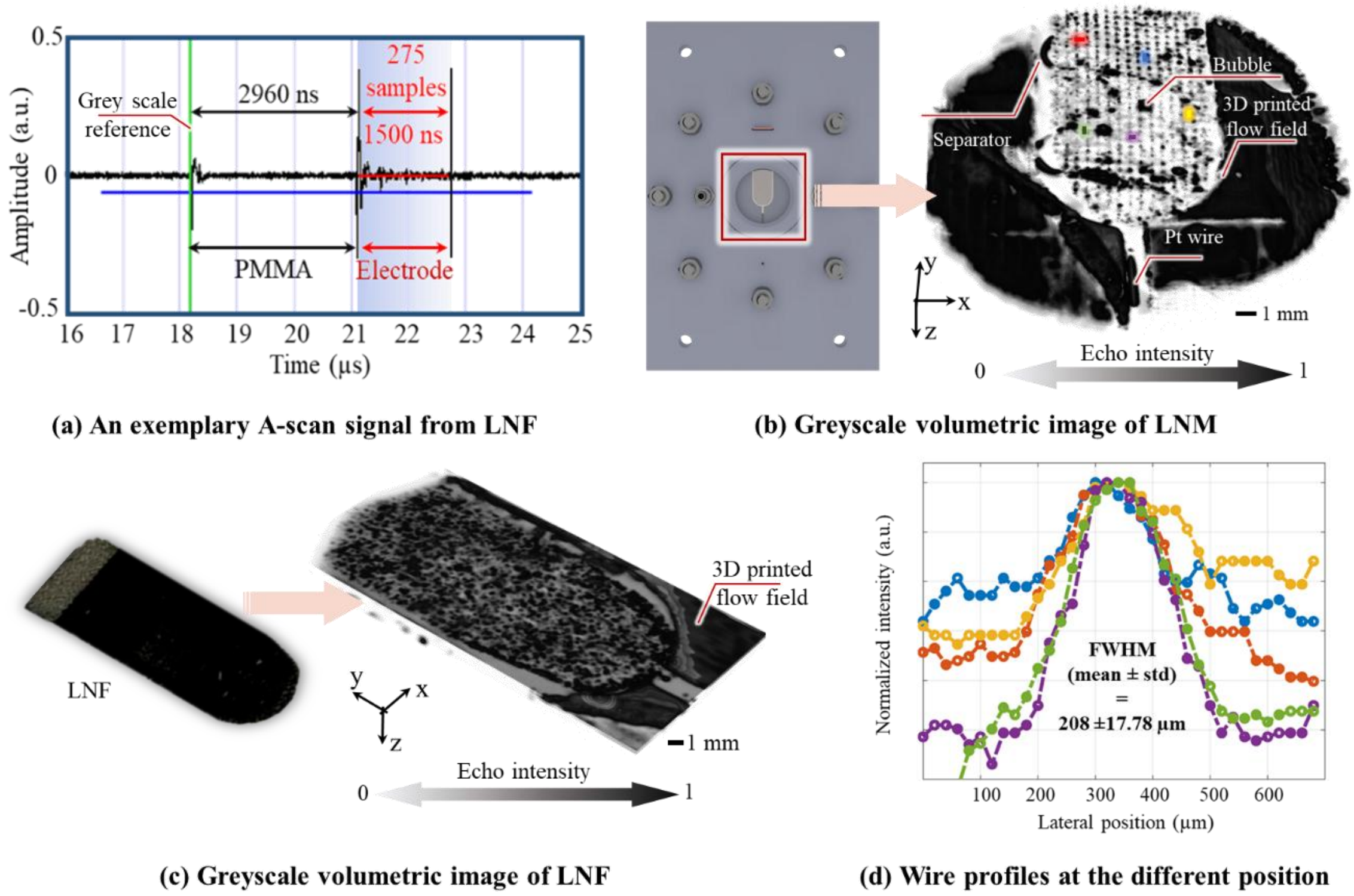

**(a) An exemplary A-scan signal from LNF**

**(b) Greyscale volumetric image of LNM**

**(c) Greyscale volumetric image of LNF**

**(d) Wire profiles at the different position**

**Fig. 5. Characterizations of volumetric SAM Imaging performance.** (a) An exemplary A-scan signal used for volumetric image reconstruction where the selected time window and reference point are indicated. (b) The volumetric image of the Nickel Mesh (LNM) in greyscale where the colour lines indicate the positions for lateral spatial resolution evaluation. (c) The volumetric image of the Nickel Foam (LNF) in greyscale. (d) The wire profiles extracted at the corresponding positions marked in (b).

A total of 275 sampling points were selected within the time window for reconstruction [**see Fig. 5 (a)**]. This resulted in a voxel size in the axial ($z$) direction of

$$L_{vox} = {}^{d}/_{n_{samples}} = {}^{1.185\ mm}/_{275} = 4.3\ \mu m \quad (7).$$

Accordingly, the volume of each voxel is calculated as:

$$vol_{vox} = 20 \times 20 \times 4.3\ \mu m^3 \quad (8).$$

The well-defined geometry of the woven mesh grid on the LNM electrode provided a reliable basis for evaluating the actual lateral resolution realized by the SAM system. **Fig. 5 (d)** plots the wire profiles extracted at various positions on the LNM grid, as indicated by the colour dots in **Fig. 5 (b)**. The mean and standard deviations of the full-width at half-maximum (FWHM) of these profiles were determined as 208 ± 17.78 μm. Given that the actual wire diameter on the LNM was 140 μm, this implies an average broadening of approx. 68 μm in the SAM images. This broadening corresponds to the FWHM of the system's point spread function (PSF) under in the experimental environment, and agrees well with the theoretical lateral resolution of approx. 70 μm, as modelled in **Fig. 4 (b)**. In other words, any individual bubbles separated by more than 70 μm can be resolved in the acquired volumetric images.

### 2.4 Pulse electrolysis strategy for time-resolved imaging

Given the relatively long image acquisition time of 180 s per volumetric frame, a pulse electrolysis scheme was employed to simulate time-resolved imaging of gas-liquid transport during HER, as shown in **Fig. 6**. This scheme consisted of six repeated on-off cycles, each comprising an on-period lasting 0.5 s, during which a constant current corresponding to a current density of - 80 mA/cm$^2$ normalized to the surface area of the electrodes was applied for electrolysis. Each on-period was followed by a 300 s off-period allowing for image acquisition.

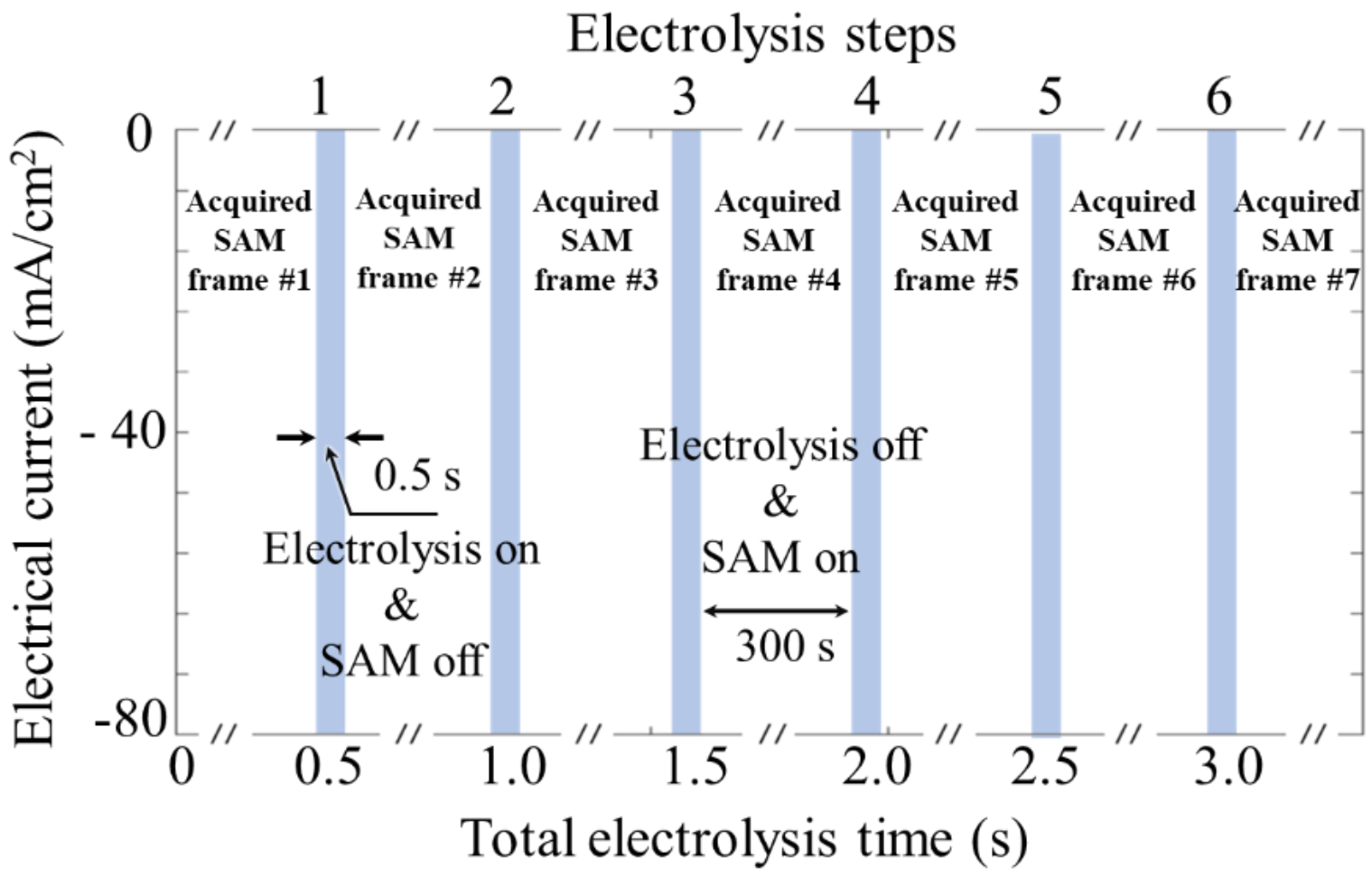

**Fig. 6. Timing of the pulse electrolysis scheme used to mimic time-resolved SAM imaging of fluid transport during electrolysis.**

Although each volumetric X-scan required only 180 s, the off-period was extended to 300 s to accommodate manual operation of the SAM interface, as the SAM system was not synchronized with the electrochemical workstation. Including the initial pre-electrolysis state after cell conditioning, seven volumetric frames were acquired for each electrode, offering a first proof-of-principle demonstration of the SAM imaging strategy. Since no additional flow was applied throughout the pulse electrolysis, it was assumed that gas bubbles generated during the on-period remained stationary (i.e., neither dissolved nor displaced) during the subsequent off-period, ensuring faithful capture of their spatial distribution.

## 3. Results and discussions

With the validated volumetric SAM imaging performance, we apply this technique to investigate the evolution of hydrogen bubbles on the Ni woven mesh and open-cell foam electrodes. This section presents the *operando* imaging results and discusses the observed bubble behaviours. To enable quantitative analysis, tailored image segmentation strategies were developed for each electrode type, adapting with their distinct structural and acoustic characteristics.

### 3.1 Bubble evolution on woven mesh electrodes

According to the acoustic impedance values listed in **Table I**, a nearly total acoustic reflection at the bubble/electrolyte interface can be expected. In comparison, the reflection coefficient of acoustic energy at the Ni wire/electrolyte interface is calculated to be approx. 88%. As shown in **Fig. 7 (a)**, gas bubbles on the Nickel Mesh (LNM) electrode are characterized by higher echo intensities, which appear as darker greyscale values in the SAM images. Moreover, the regular woven mesh geometry ensures a uniform ultrasound incidence over the whole electrode surface, resulting in spatially consistent echo intensities from both bubbles and Ni wires [see inset of **Fig. 7 (a)**]. Such distinct echo intensity contrast allows gas bubbles to be effectively and reliably segmented from woven mesh electrodes using a direct thresholding method based on a single global intensity threshold.

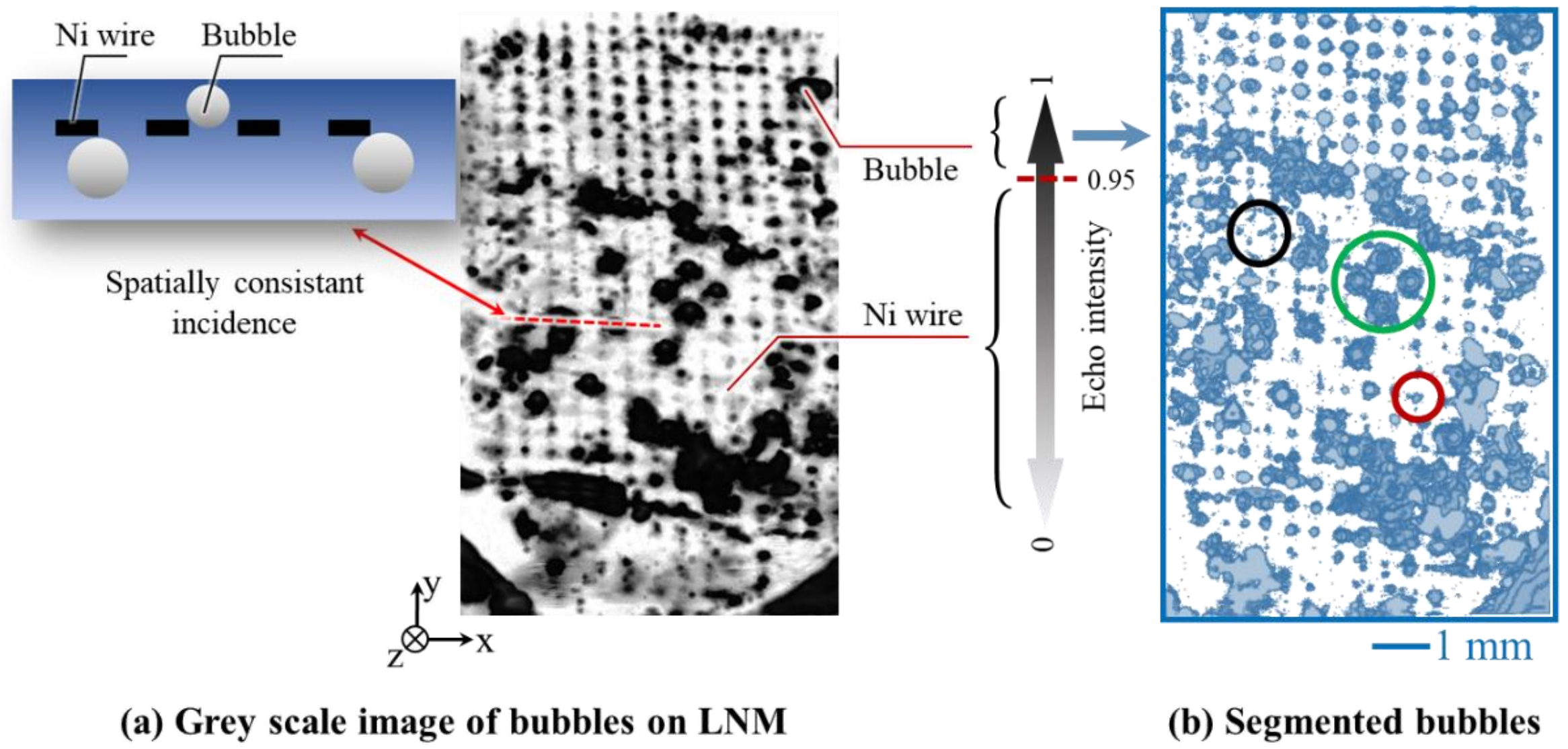

**Fig. 7. Direct thresholding-based segmentation of gas bubbles on woven mesh electrodes scheme.** (a) A grey scale SAM image of the LNF, the inset schematically illustrates the spatially consistent echo intensities. (b) The top view of the segmented bubbles corresponding to (a).

To implement this direct thresholding segmentation scheme, each volumetric image of the LNM was first smoothed using a median filter to supress salt and pepper noise. A global threshold value was then empirically defined as 95 % of the highest echo intensity within the electrode region, for all the acquired LNM images. This means that the voxels with echo intensity values above the threshold value were designated as gas bubbles, as shown in **Fig. 7**. The top view of the resulting 3D binary image is shown in **Fig. 7 (b)**, in which distinct bubble morphologies are visible, including clusters of small bubbles on wire surfaces (black circle), localized bubbles at wire intersection points (red circle), and larger bubbles trapped within the mesh pores (green circle).

The spatiotemporal distributions of these produced hydrogen bubbles on the LNM are shown in **Fig. 8 (a)**, in form of 3D binary images. It is observed that, the small bubbles initially appear across the wire surfaces and at wire intersections, forming distinct clusters. As electrolysis proceeds, larger bubbles become visible, particularly those trapped within the pores of the woven mesh. Notably, the gas distribution is not uniform: the central region of the LNM remains relatively bubble-free compared to the more densely populated outer zones. This likely aroused from non-uniform current density distribution across the electrode, which is influenced by the geometry of the ring-shaped CE, as shown in **Fig. 2 (c)**.

This direct thresholding approach enables quantifications of the volume of bubbles that are attached to the LNM from the SAM images. For this, the total number of voxels ($n$) identified as gas in the segmented binary volumetric images within the LNM occupied region are counted. Since each voxel has a known volume, the gas volume can be calculated using (9), in which the summation is performed over all voxels classified as gas.

$$V_{gas} = vol_{vox} \cdot \sum n|_{\ vox=Gas} \tag{9}$$

**Fig. 8 (b)** plots the volumes of the gas bubbles on the LNM and the absolute value of its potential over in the total 3 s pulse electrolysis carried out at a geometrical current density of - 80 mA/cm$^2$. It is observed that the gas volume and the potential at the LNM both increase rapidly, indicating the gas accumulation-induced overpotential, due to increased bubble surface coverage and ohmic resistance of the bubble layer.

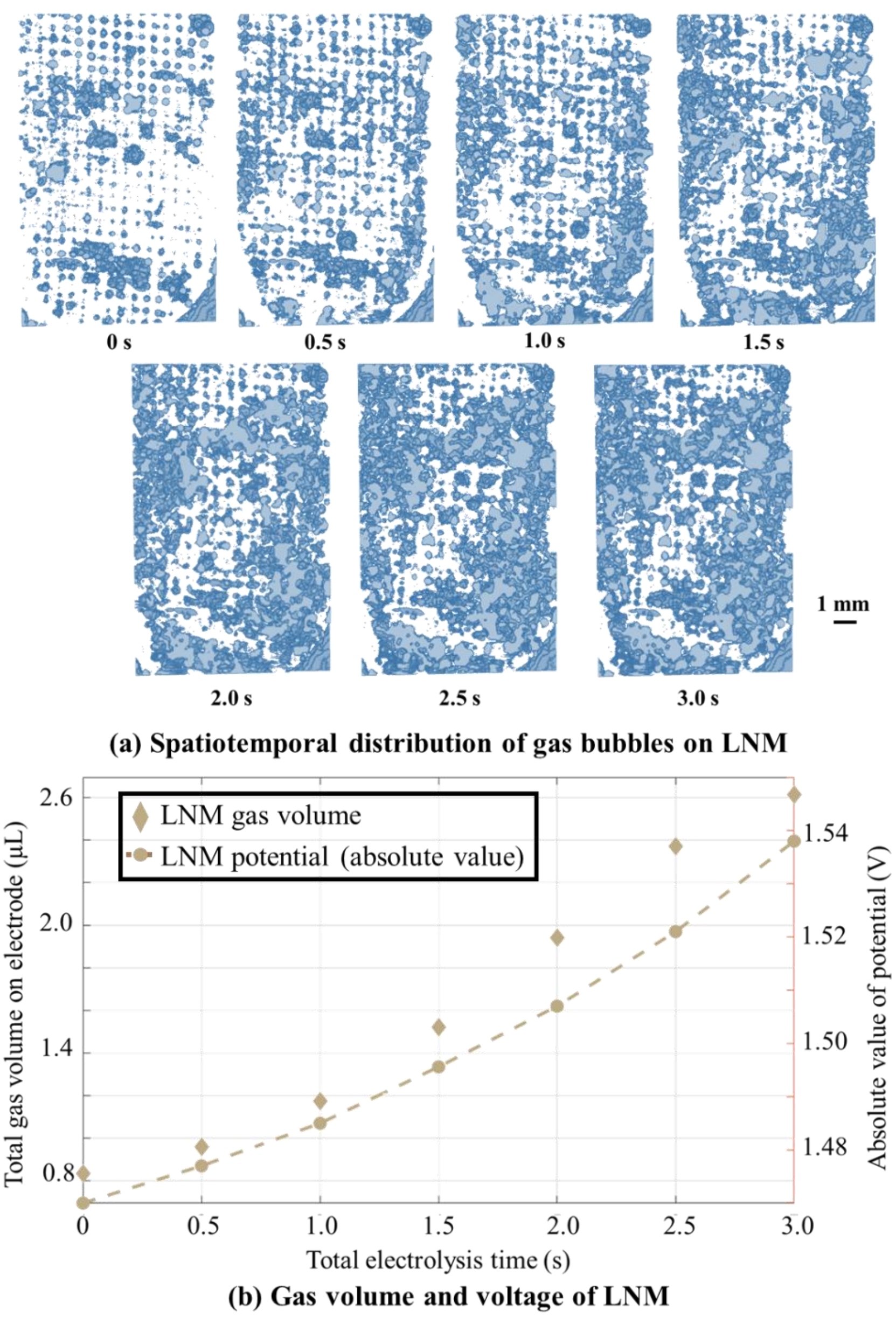

**Fig. 8. Quantification of gas volume on the Nickel Mesh (LNM) electrode operated at a geometrical current density of 80 mA/cm$^2$.** (a) The spatiotemporal distribution of hydrogen bubbles on the LNM. (b) The volume of bubbles attached to the LNM and the and absolute potential value at the LNM over time.

### 3.2 Bubble evolution in open cell foam electrodes

In contrast to the woven meshes with simple structures, open cell foam electrodes exhibit a stochastic, complex, and highly interconnected 3D architecture consisting of randomly oriented struts and irregular pore geometries. The interaction of ultrasound with the foam is highly dependent on the local geometry, due to the diverse patterns of reflection, scattering, and transmission at each scanning point. This leads to significant variations in the local incident acoustic energy, which in turn results in significant fluctuations in the echo intensities from both the foam struts and the gas bubbles throughout the 3D imaging volume. Consequently, such highly heterogeneous and spatially inconsistent acoustic responses make the direct global thresholding approach unsuitable for accurate segmentation of gas contents in open cell foam electrodes.

To address this challenge, a differential image subtraction scheme is used in this study. While the gas content evolves dynamically, the open cell foam structure remains static and time-invariant during electrochemical operation. Furthermore, it is reasonable to assume that the gas content in the electrode before electrolysis is minimal. Thus, during electrolysis, hydrogen bubbles formed within the foam pores displace the electrolyte, thereby increasing the local echo intensity in subsequent SAM acquisitions. This enables a relative quantification approach: the increase in local echo intensity between two volumetric images acquired before and after electrolysis reflects bubble growth.

To implement this subtraction-based segmentation method, the greyscale volumetric SAM image of the Nickel Foam (LNF) acquired before electrolysis [marked with a red star in **Fig. 9 (b)**] was first processed using a median filter to suppress noise, and the image was then binarized to identify the pore network of the LNF. An example of 3D binary image of the LNF before electrolysis is shown in **Fig. 9 (b)** (black star). Here, the binarization was carried out with the MATLAB's default implementation of Otsu's method [45]. From this initial binary image, the total pore volume i.e., the volume where bubbles may appear, is calculated by summing the number of voxels identified as pore:

$$V_{pore} = \sum n\Big|_{vox=pore}^{initial_image} \tag{10}$$

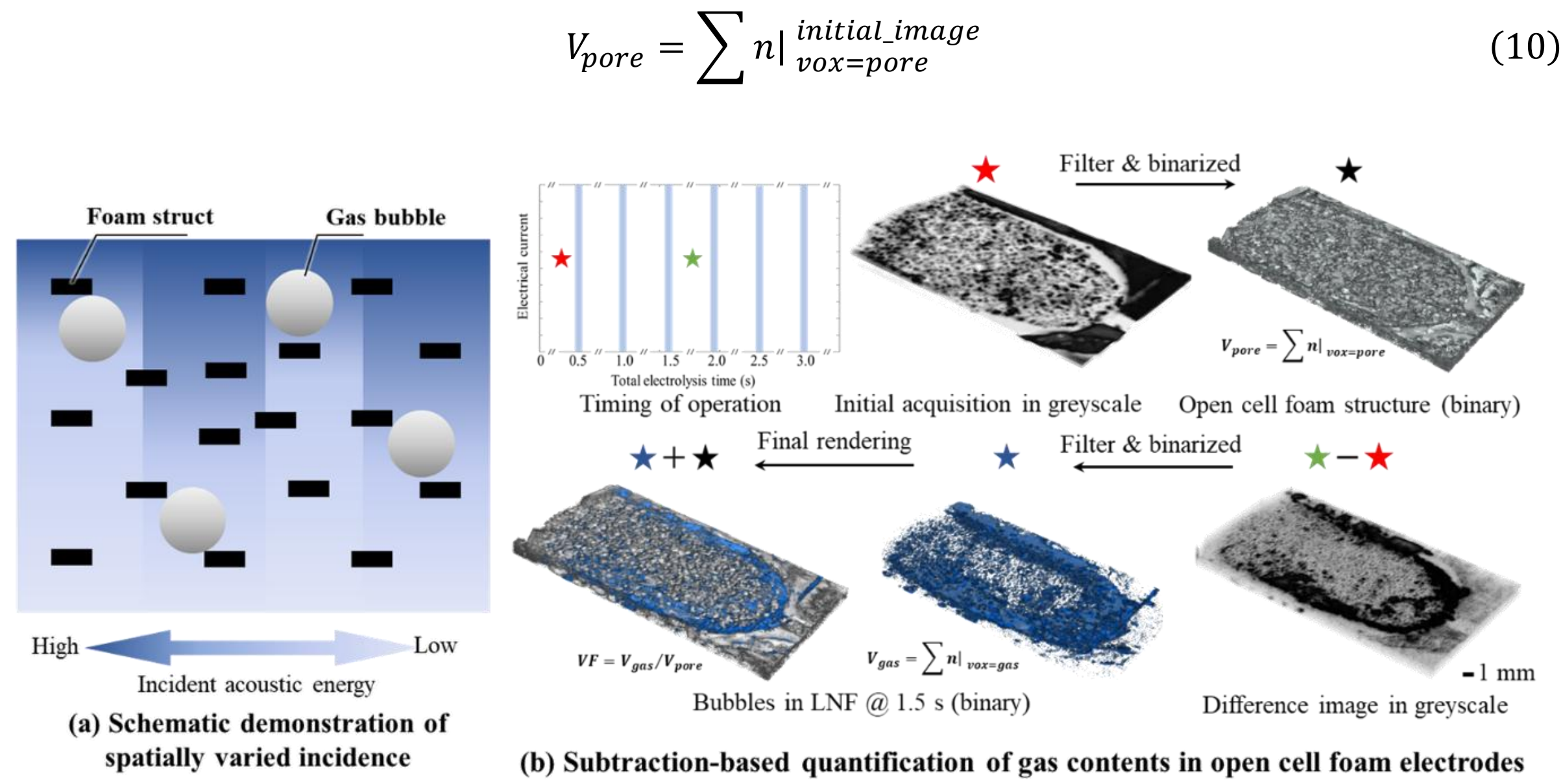

**Fig. 9. Differential image subtraction-based segmentation of gas contents in open cell foam electrode scheme.** (a) Schematic demonstration of spatially varied incident acoustic energy into the open cell foam electrode. (b) Graphical representation of the subtraction-based segmentation and quantification scheme.

To quantify the hydrogen volume produced during electrolysis, a difference image is first computed by subtracting the initial image from the one acquired after a given electrolysis step. This differential image is processed using the same median filtering and binarization procedure. The resulting binary image isolates regions of increased echo intensity, which are attributed to newly evolved gas contents. An example of this processing pipeline is shown in the lower row of **Fig. 9 (b)**, demonstrating hydrogen evolution in the LNF after 1.5 s of electrolysis. The gas content in such open cell foam electrodes is quantified using the volume fraction (VF) of pores occupied by gas, calculated as:

$$VF = {V_{gas}}/{V_{pore}} \tag{11}$$

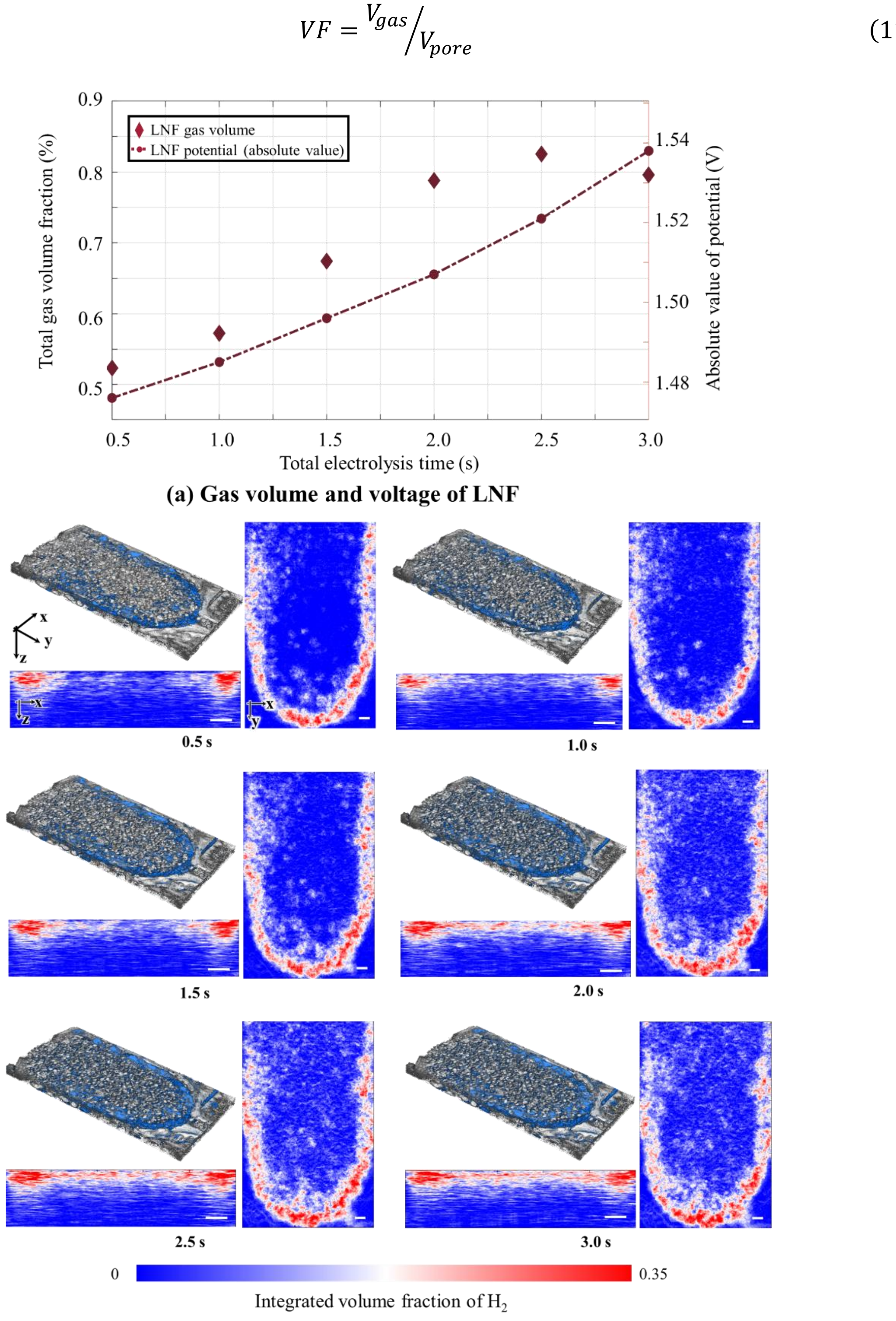

**Fig. 10. Quantification of gas contents in the Nickel Foam (LNF) electrode operated at a geometrical current density of 80 mA/cm$^2$.** (a) The gas volume and absolute value of potential of the LNF. (b) The spatiotemporal distribution of hydrogen volume fraction in the LNF where the scale bars are 1 mm.

With this approach, the gas content in the LNF during electrolysis at a geometrical current density of - 80 mA/cm$^2$ is quantified. **Fig. 10 (a)** plots the volume fractions of gas in the LNF and the absolute value of its potential over time, in which a direct correlation is observed. Moreover, the spatial bubble distributions in **Fig. 10 (b)** reveal a distinct temporal hydrogen evolution pattern. Initially, the hydrogen evolution predominantly occurs at the outer edges of the LNF, forming large bubble clusters. This edge favoured evolution and accumulation behaviour likely resulted from greater local current densities due to the ring-shaped CE. As electrolysis proceeded beyond 1.5 s, the outer pores become increasingly saturated with gaseous $H_2$, and bubble formation begins to emerge more prominently in the central regions of the LNF. These interior bubbles are smaller and more sparsely distributed, forming at discrete nucleation sites.

### 3.3 Applicability and limits of SAM imaging

In this study, the feasibility of using high-resolution volumetric SAM imaging to quantify gas evolution in technically relevant porous Ni electrodes of operating AWE was experimentally demonstrated. Compared with conventional optical or ultrasound imaging in the low MHz frequencies, SAM imaging using high frequency focused ultrasound offers a unique combination of high spatial resolution, volumetric imaging capability, and strong sensitivity to local gas contents in opaque materials.

Despite these advantages, the temporal resolution of SAM imaging towards real-time volumetric acquisitions must be justified. In this present study, a temporal resolution of 180 s was achieved for scanning an area of 900 mm², which limits its ability to resolve fast bubble dynamics in real-time. This limitation arises primarily from the time-consuming pointwise mechanical scanning required to cover large imaging volumes. However, higher temporal resolution is feasible within smaller RoI. For instance, with the current scanning speed, a temporal resolution of 0.5 s can be achieved for an area of 2.5 mm².

A comparison with state-of-the-art synchrotron X-CT systems helps to further position SAM instrumentation within modern *operando* imaging based analytical tools. S. De Angelis *et al.* recently reported visualization of bubble transport in a customized PEMWE test cell using one of the most cutting-edge synchrotron X-CT system, in which a temporal resolution of 1.4 s within a RoI area of 3.3 mm × 4.6 mm was achieved [27]. When normalized to the same imaging area, the SAM imaging in this work demonstrates a comparable acquisition speed of 3 s per frame, highlighting its competitiveness in terms of resolution-scalability balance. Moreover, the sequential scanning nature of SAM imaging is inherently compatible with further acceleration strategies towards imaging in larger volumes. These include incorporating multiple simultaneous scanning transducers or implementing computational ultrasound imaging with customized transducer arrays to increase the spatial bandwidth product of the imaging aperture [46], [47]. However, expanding the FoV of CT systems typically leads to a steep trade-off between the temporal resolution and imaging quality.

Beyond the technical scalability, SAM imaging also offers practical advantages in terms of accessibility and cost. In contrast to synchrotron X-CT systems that require access to beamlines and involve complex radiation safety protocols, SAM instrumentations can be operated in standard laboratory environments with relatively low infrastructure demands. This makes SAM a highly attractive and cost-effective alternative for routine *operando* studies of fluid transport in gas evolving electrochemical systems, especially when diverse designs need to be investigated.

## 4. Conclusions and outlook

This work experimentally demonstrates the feasibility of high-resolution volumetric SAM imaging for quantifying gas evolution in porous Ni electrodes of AWE under *operando* conditions. These contribution opens new horizon of using SAM instrumentations as an effective and reliable analytical tool for investigating fluid transport in a wide range of electrochemically gas evolving systems.

Based on this methodological foundation, future works will improve the current instrumentation towards real-time imaging, and apply it to study gas/liquid transport under more practically relevant operating scenarios, in order to systematically investigate the complicated interplay among electrode configurations, flow field designs, cell operating conditions and bubble removal performance. The resulting insights are expected to support design-driven optimization of large-scale AWE systems.

## Acknowledgements

The authors would like to thank L. Buettner, R. Nauber and H. Emmerich for valuable discussions, and B. Birki for the assistance on SAM experiments. The authors specially thank T. Djuric-Rissner for coordinating the collaborations between TU Dresden and PVA TePla Analytical Systems GmbH. This project is supported with funds provided by the Deutsche Forschungsgemeinschaft (DFG) under grant 512483557 (BU2241/9-1), the Helmholtz Association Innovation pool project "Solar Hydrogen" and the Hydrogen Lab of the School of Engineering of TU Dresden.

## Author contributions

Z.D.: Conceptualization, formal analysis, investigation, methodology, validation, visualization, writing & editing. H.R.: Conceptualization, investigation, methodology, review & editing. Z.R.: Investigation & review. R.B.: Investigation & review. R.R.: Investigation. X.Y.: Conceptualization, project administration, review & editing. P.C.: Resources & review. A.F.L.: Resources. K.E.: Funding acquisition, review & editing. J.C.: Supervision & review. D.W.: Conceptualization, project administration, review & editing.

## Declaration of interests

The authors declare no competing interests.

## Supplementary Information

### Electrodes preparations

The Ni woven mesh sheets and open cell foam used in this work were obtained from HAVER & BOECKER and Fraunhofer IFAM Dresden, respectively.

The laser texturing process was carried out using a picosecond slab-shaped solid-state laser (Edgewave InnoSlab, Würselen, Germany) emitting at a wavelength $\lambda$ of 532 nm and with a pulse width of 12 ps. The laser beam is guided by high reflective mirrors towards a beam expander and further to a 2D-galvo scanner (Raylase Super Scan III, Weßling, Germany). Furthermore, a f-theta lens is applied with a focal length of 100 mm, leading to a beam diameter in the focal plane of 18 µm. The scanning velocity was set to 6.25 mm/s with a pulse repetition of 10 kHz and a pulse energy of 100µJ. The texturing process was performed using a unidirectional scan strategy. Cross-like and line-like patterns were generated on the LNM and LNF, respectively. The spacing between of each generated structure was set to 40 µm for all investigated specimen.

### Electrode contact angle characterization

The Wilhelmy plate method was applied to measure the apparent contact angle of the wicking state electrodes. For this, the measured electrodes were immersed into the electrolyte i.e., 8 wt% KOH solution. Multiple images of each electrode were captured by a digital camera (Pentax K1) equipped with a macro lens. The contact angles were then obtained in ImageJ.

### Surface morphology characterization of open cell foam electrodes

The surface morphologies of the four electrodes were acquired using a scanning electron microscopy (Zeiss 1540 XB) with a 20 kV accelerating voltage.